\documentclass[conference]{IEEEtran}
\IEEEoverridecommandlockouts
\usepackage{tikz}
\usepackage{pgfplots} 
\usepackage{pgfgantt}
\usepackage{pdflscape}
\pgfplotsset{compat=newest} 
\pgfplotsset{plot coordinates/math parser=false}
\usepackage{amsmath}
\usepackage{mathtools}
\usepackage{bm}
\usepackage{adjustbox}
\usepackage{acronym}
\acrodef{OFO}{Online Feedback Optimization}
\acrodef{OPF}{Optimal Power Flow}
\usepackage{float}
\usepackage[hyperindex=true, %
  pdftitle={Real-time Curative Actions for Power Systems via Online Feedback Optimization},%
  pdfauthor={Lukas Ortmann, Gianni Hotz, Saverio Bolognani, Florian D\"{o}rfler},%
  colorlinks=true,%
  pagebackref=false,%
  plainpages=false,%
  pdfpagelabels,%
  linkcolor=black,%
  citecolor=black,%
  filecolor=black,%
  urlcolor=black%
  ]{hyperref}
 \usepackage{cite}

\title{Real-time Curative Actions for Power Systems\\via Online Feedback Optimization}

\author{Lukas Ortmann, Gianni Hotz, Saverio Bolognani, Florian D\"{o}rfler% <-this % stops a space
\thanks{The authors are with the Automatic Control Laboratory, ETH Z\"urich, Physikstrasse 3, 8092 Z\"urich, Switzerland. This research has been supported by ETH Zürich funds}%
\thanks{Email: \, {\tt\footnotesize \{ortmannl,bsaverio,doerfler\}@ethz.ch}}}

\begin{document}

\maketitle

\begin{abstract}
Curative or remedial actions are the set of immediate actions intended to bring the power grid to a safe operating point after a contingency. The effectiveness of these actions is essential to guarantee curative $N-1$ security. Nowadays, curative actions are derived ahead of time, based on the anticipated future grid state. Due to the shift from steady to volatile energy resources, the grid state will frequently change and the curative actions would need to be pre-planned increasingly often. 
Furthermore, with the shift from large bulk production to many small decentralized energy sources more devices need to be actuated simultaneously to achieve the same outcome.
Instead of pre-planning, we propose to calculate these complex curative actions in real-time after the occurrence of a contingency. 
We show how the method of \ac{OFO} is well suited for this task.
As a preliminary demonstration of these capabilities, we use an \ac{OFO} controller, that after a fault, reduces the voltage difference over a breaker to enable the operators to reclose it. This test case is inspired by the 2003 Swiss-Italian blackout, which was caused by a relatively minor incident followed by ineffective curative actions.
Finally, we identify and discuss some open questions, including closed-loop stability and robustness to model mismatch.
\end{abstract}

\begin{IEEEkeywords}
Online Feedback Optimization, Emergency Power System Operations, Curative Actions, Real-time Control, Curative $N-1$ Security
\end{IEEEkeywords}

\section{Introduction}

The electrical power grid is a critical infrastructure and the backbone of modern society. Its uninterrupted operation is crucial and it is essential that security can be guaranteed at all times even when contingencies occur, i.e., a transformer, power plant, or power line is disconnected. Therefore, the grid is operated following the $N-1$ criterion, meaning that the grid must be in a safe state even if any single element fails. This is also referred to as \emph{preventive} $N-1$ security. The ENTSO-E grid code used in Europe allows temporary overloads in case of a contingency if curative or remedial actions are defined upfront to bring the system back to a safe operating point \cite[Article 32(2)]{grid_code}. The North American Electric Reliability Cooperation (NERC) allows for a \emph{Remedial Action Scheme} that automatically takes corrective actions \cite{NERC}.
Permitting such temporary violations relaxes the $N-1$ criterion to \emph{curative} $N-1$ security and enlarges the range of allowed grid configurations, which enables more economical grid operation \cite{westermann2019curative, monticelli1987security}. This idea of using curative actions dates back at least to the 1980s~\cite{monticelli1987security}, where security-constraint \ac{OPF} with curative actions were proposed.
Using curative $N-1$ security is an active field of research by, e.g., German Transmission Grid Operators and universities as it helps to utilize the grid to a larger extent~\cite[Subsection 5.2]{Netzentwicklungsplan2030}.
Available curative actions are, e.g., changes of active power generation setpoints, operating points of high-voltage direct current systems, voltage set-points or reactive power injections, and tap changers positions of phase-shifting transformers. Lately, the shift to decentralized generation also enables distribution grids to provide curative actions~\cite{SiemensPSCC2022}. An overview of curative actions is presented in \cite[Table III]{yamashita2009analysis}.

Currently, the curative actions are decided manually by operators based on long-term experience or based on a library of case studies created by solving \ac{OPF}s for a set of contingencies so that the actions are available in case they occur. As the share of production from volatile and unpredictable renewable energy sources increases, the grid is expected to operate at different operating points throughout the day, which requires operators to update their curative action plan more often than today. Moreover, determining the best emergency response is already complex nowadays, but it will become more complex in the future because large power plants are being replaced by decentralized energy resources and therefore the number of actuators needed for effective curative actions will increase. Overall, operators will need to determine curative actions more often, and those actions will be more complicated.
Last but not least, when a contingency occurs, the operating personnel needs to implement the curative actions quickly while guaranteeing that those actions will not lead to new problems elsewhere. 

In contrast to the current practice, we propose to employ a closed-loop control scheme to derive curative actions in real-time after the occurrence of a contingency.
This has the following advantages: 1) the current operating point of the grid is taken into account; 2) the feedback nature of closed-loop control provides robustness to model mismatch; 3) due to the low computational complexity, the curative actions are promptly implemented to quickly drive the grid to a feasible operating point.

The control strategy that we propose is based on \ac{OFO}, a methodology that allows converting iterative optimization algorithms into real-time robust feedback controllers \cite{hauswirth2021optimization, bernstein2019online, lawrence2020linear, colombino2019online,bianchin2021time}. These controllers can then be used to drive a system to the optimum of a constrained optimization problem, which, in the application that we are considering, defines the safe operating region of the grid. Such controllers do not need a full model of the system and guarantee constraint satisfaction even in the presence of model mismatch. 
There exist several different versions, e.g., distributed, centralized, model-based, and model-free controllers \cite{he2022model,picallo2021adaptive}. 
They are well suited for several real-time optimization problems in power systems \cite[Section IV]{dan_review}, and they have also been experimentally validated \cite{ortmann2020experimental, ortmann2020fully}.

To show how an \ac{OFO} controller could help control the grid during an emergency power system operation, we take inspiration from the 2003 Swiss-Italian blackout. In that blackout, a breaker could not be reclosed because of the excessive voltage angle difference across the breaker. Using the IEEE~39~bus model we set up a grid in which opening a breaker leads to a high angle difference, which we then reduce using an \ac{OFO} controller.

The structure of the paper is as follows. In Section~\ref{sec:blackout} we describe the Swiss-Italian blackout and in Section~\ref{sec:simulation_setup} we present the simulation setup we are using to reconstruct the underlying problem of this blackout. Afterward, in Section~\ref{sec:OFO}, we design an \ac{OFO} controller that determines effective curative actions in real-time. We present the results of our simulations in Section~\ref{sec:results} and conclude the paper in Section~\ref{sec:conclusion}.

\section{An example of unsuccessful emergency operations: the 2003 Swiss-Italian Blackout}
\label{sec:blackout}
On September 28th, 2003, Italy was importing a large amount of power from its neighboring countries. At 3:01, a 380~kV line in Switzerland tripped due to a tree flashover. Due to the large power flow toward Italy, there was a high phase angle difference of $42^\circ$ across the now open circuit breaker. Reclosing this breaker would have resulted in high transient stress for generators located in that region and therefore a local protection system prevented the operators to reclose the line as long as the angle difference was larger than $30^\circ$. Meanwhile, because of the open line, power flow increased on other lines, leading to one of them operating at 110\% of its capacity. This resulting overload still satisfied the curative $N-1$ criterion, assuming that it could be promptly mitigated. The Swiss operators deployed several control actions to enable reclosing the breaker and to lower the overloading, but did not succeed. The line overheated, which resulted in excessive sag of the conductor. At 3:25, after 24 minutes, a tree flashover occurred and the line was automatically disconnected. The remaining power lines immediately overloaded and were disconnected, leading to the largest Italian blackout in history \cite{UCTE_report}. The estimated cost of this 18-hour blackout is 1.2~billion Euros~\cite{price_blackout}.

\section{Simulation Setup}
\label{sec:simulation_setup}
\begin{figure}[tb]
    \centering
	\includegraphics[width=0.8\columnwidth]{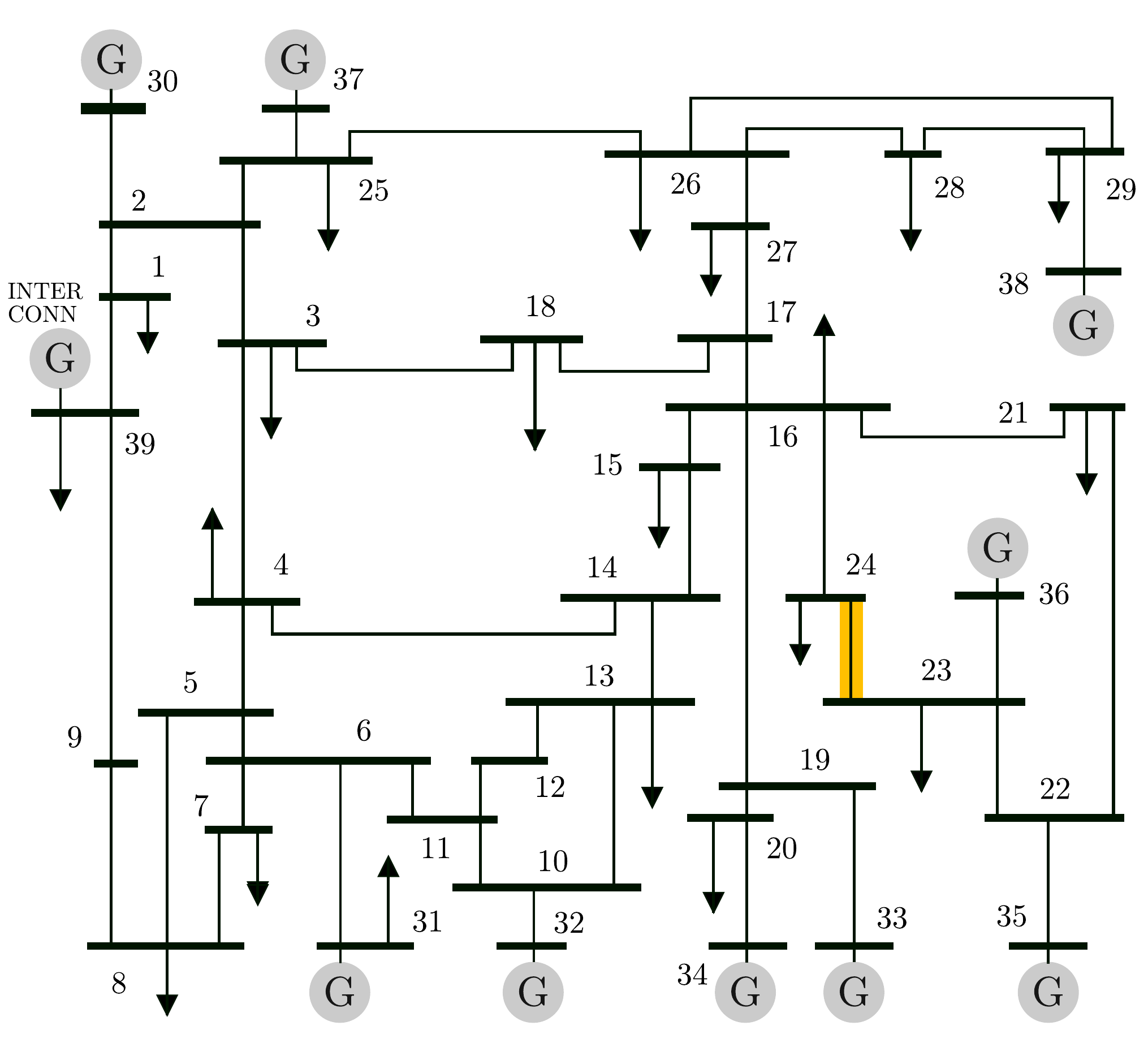}
		\caption{IEEE39 test case. The orange line is tripped and needs to be reclosed to bring the system back to a safe state.}
	\label{fig:IEEE39}
\end{figure}

\begin{figure*}[]
    \centering
    \begin{tikzpicture}[scale=0.9,every node/.style={transform shape}]
        \draw[-stealth] (-1,4) to (-1,0.5) to (-0.5,0.5) node[above] {$\omega$} to (0,0.5);
        \draw (0,0) rectangle +(1.5,1) node[pos=.5] {GOV};
        \draw[-stealth] (7,3) to (7.5,3) node[above] {$\omega$} to (8,3) to (8,5) to (-1,5) to (-1,4) to (-0.5,4) node[above] {$\omega$} to (0,4);
        \draw (0,3.5) rectangle +(1.5,1) node[pos=.5] {PSS};
        \draw[-stealth] (1.5,4) to (2,4) node[above] {$v_{pss}$} to (2.5,4);
        \draw (2.5,2) rectangle +(1.5,2.5) node[pos=.5] {AVR};
        
        \draw (0.75,2.5) circle [radius=0.25] node {$+$};
        \draw[-stealth] (-1.75,1.75) to (-0.5,1.75) node[above] {$v$} to (0.75,1.75) to (0.75,2.25) node[right] {$-$};
        \draw[-stealth] (-2,3) to (-0.5,3) node[above] {$v_{OFO}$} to (0.75,3) to (0.75,2.75);
        \draw[-stealth] (1,2.5) to (2,2.5) node[above] {$\Delta v$} to (2.5,2.5);
        
        \draw (5.5,1.0) rectangle +(1.5,3) node[pos=.5] {GEN};
        \draw[-stealth] (4,3.5) to (5,3.5) node[above] {$E_f$} to (5.5,3.5);
        \draw[-stealth] (0.75,1.75) to (4.25,1.75) to (4.25,2.5) to (5,2.5) node[above] {$v$} to (5.5,2.5);
        \draw[-stealth] (4,0.5) to (4.75,0.5) to (4.75,1.5) to (5,1.5) node[above] {$p_m$} to (5.5,1.5);
        \draw[-stealth] (7,2) to (7.5,2) node[above] {$i_{inj}$} to (9.5,2) to (9.5,1.5) to (10,1.5) node[above] {$\bm i_{inj}$} to (10.5,1.5);
        
        \draw (3.75,0.5) circle [radius=0.25] node {$+$};
        \draw[-stealth] (1.5,0.5) to (2,0.5) node[above] {$p_{GOV}$} to (2.5,0.5) to (2.5,1) to (3.75,1) to (3.75,0.75);
        \draw[-stealth] (2.75,-1.5) to (2.75,0) node[left] {$p_{OFO}$} to (2.75,0.5) to (3.5,0.5);
        \draw[-stealth] (3.75,-2) to (3.75,0) node[right] {$p_{AGC}$} to (3.75,0.25);
        % Frame
        \draw[dotted] (-1.25,-0.25) to (6.75,-0.25) node[above] {\emph{Generating unit}} to (8.25,-0.25) to (8.25,5.25) to (-1.25,5.25) to (-1.25,-0.25);
        \draw[dotted] (-0.75,-0.25) to (-0.75,-0.75) to (8.75,-0.75) to (8.75,4.75) to (8.25,4.75);
        % Outputs
        \draw[dotted] (8.5,1.5) to (9.5,1.5);
        \draw[dotted] (8.5,2.5) to (9.5,2.5) to (9.5,3);
        \draw[-stealth] (8,3) to (10,3) node[above] {$\bm\omega$} to (10.5,3);
        % Inputs
        \draw[dotted] (-2,2.5) to (-1.25,2.5);
        \draw[dotted] (-1.75,1.25) to (-1.25,1.25);
        \draw[dotted] (3.25,-1.5) to (3.25,-0.5);
        \draw[dotted] (4.25,-2) to (4.25,-0.5);
        % PF, OFO
        \draw (10.5,1) rectangle +(1.5,1) node[pos=.5] {PF};
        \draw[-stealth] (12,1.5) to (12.5,1.5) node[above] {$\bm u$} to (13,1.5);
        \draw (12.5,1.5) to (12.5,-1.25) to (-1.75,-1.25) to (-1.75,1.75);
        \draw (13,0) rectangle +(1.5,2) node[pos=.5] {OFO};
        \draw (14.5,0.5) to (15,0.5) node[above] {$\bm p_{OFO}$} to (15.5,0.5) to (15.5,-1.5) to (2.75,-1.5);
        \draw (14.5,1.5) to (15,1.5) node[above] {$\bm v_{OFO}$} to (15.75,1.5) to (15.75,-1.75) to (-2,-1.75)  to (-2,3);
        % AGC
        \draw (10.5,2.5) rectangle +(1.5,1) node[pos=.5] {AGC};
        \draw (12,3) to (12.5,3) node[above] {$\bm p_{AGC}$} to (16,3) to (16,-2) to (3.75,-2);
        % Dots
        \filldraw[black] (0.75,1.75) circle (1pt);
        \filldraw[black] (-2,2.5) circle (1pt);
        \filldraw[black] (-1.75,1.25) circle (1pt);
        \filldraw[black] (-1,4) circle (1pt);
        \filldraw[black] (8,3) circle (1pt);
        \filldraw[black] (9.5,1.5) circle (1pt);
        \filldraw[black] (9.5,3) circle (1pt);
        \filldraw[black] (12.5,1.5) circle (1pt);
        \filldraw[black] (3.25,-1.5) circle (1pt);
        \filldraw[black] (4.25,-2) circle (1pt);
        
    \end{tikzpicture}
    %\end{adjustbox}
    \caption{Block diagram of the dynamic power system simulator \emph{DynPSSimPy} including the OFO controller. The blocks PSS, AVR, GOV, GEN, AGC, PF and OFO correspond to the power system stabilizer, the automated voltage regulator, the governor, the synchronous machine, the excitation system, the power flow equations and the OFO controller, respectively. The blocks corresponding to one generating unit are grouped by the dotted frame, and possible additional generation units are indicated by the second dotted frame.}
\label{fig:simulation_model}
\end{figure*}
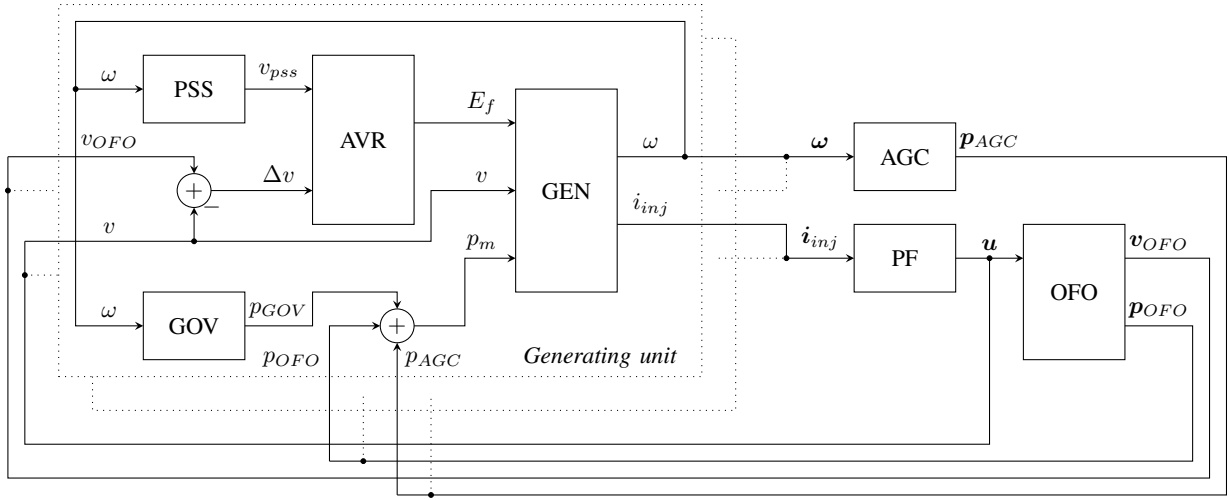

\begin{figure}[h]
    \center
    \begin{adjustbox}{width=0.9\columnwidth}
    \begin{tikzpicture}
        \draw (-0.75,0.5) to (-1.5,0.5) node[left] {$\frac{E_{f0}}{K_{ex}}$};
        \draw[stealth-] (-1,-0.25) to (-1.5,-0.25) node[left] {$\Delta v$};
        \draw (-0.75,-1) to (-1.5,-1) node[left] {$v_{pss}$};
        
        \draw (-0.75,-0.25) circle [radius=0.25] node {$+$};
        
        \draw[-stealth] (-0.75,-1) to (-0.75,-0.5);
        \draw[-stealth] (-0.75,0.5) to (-0.75,0);
        
        \draw[-stealth] (-0.5,-0.25) to (0,-0.25);
        \draw (0,-0.75) rectangle +(1.5,1) node[pos=.5] {$\frac{1+sT_a}{1+sT_b}$};
        \draw[-stealth] (1.5,-0.25) to (2,-0.25);
        \draw (2,-0.75) rectangle +(1.5,1) node[pos=.5] {$\frac{K_{ex}}{1+sT_e}$};
        \draw (2.75,-0.75) to (2.55,-0.95) to (2.35,-0.95) node[left, font=\footnotesize] {$E_{min}$};
        \draw (2.75,0.25) to (2.95,0.45) to (3.15,0.45) node[right, font=\footnotesize] {$E_{max}$};

        \draw[-stealth] (3.5,-0.25) to (4,-00.25) node[right] {$E_f$};
    \end{tikzpicture}
    \end{adjustbox}
    \caption{Block diagram of the excitation system.}
    \label{fig:excitation}
\end{figure}
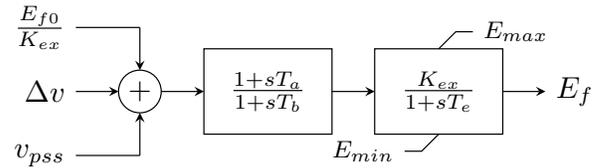

We reproduce the core phenomena of the Swiss-Italian Blackout using the publicly available IEEE 39 bus test case. It includes 10 generators, 34 lines, and 39 buses, see Figure~\ref{fig:IEEE39}.
We will trip the power line connecting buses~23 and~24, which we will not be able to reclose unless we make the voltage difference between the two buses sufficiently small.
The numerical experiment is done via the dynamic power system simulator \emph{DynPSSimPy} \cite{DynPSSimPy}, which models secondary frequency control through Automated Generation Control and the dynamics of synchronous machines, including the excitation system, power system stabilizer, and governor which includes the primary frequency control. Figure~\ref{fig:simulation_model} shows the interconnection of the different elements.
Here we give a short overview of the different components of the model.
The synchronous generators are modeled with a sixth-order system
 \begin{align*} \label{eq:ss_synch_mach}
 \begin{split}
    \Delta \dot{\omega} =& \frac{1}{2 H} \left( \frac{p_m}{\omega} - p_e - D\omega \right) \\
    \dot{\delta} =& \Delta \omega\\
    \dot{E_q'} =& \frac{E_f - E_q' - I_d (X_d - X_d')}{T_{d0}'}\\
    \dot{E_d'} =& \frac{-E_d' + I_q (X_q - X_q')}{T_{q0}'}\\
    \dot{E_q''} =& \frac{E_q'- E_q'' - I_d (X_d' - X_d'')}{T_{d0}''}\\
    \dot{E_d''} =& \frac{E_d'- E_d'' + I_q (X_q' - X_q'')}{T_{q0}''}\\
\end{split}
\end{align*}
with the speed deviation of the rotor speed from the nominal frequency $\Delta \omega$, the rotor angle $\delta$ and the internal voltages $E_q'$, $E_d'$, $E_q''$ and $E_d''$. The variables $H$, $D$, $X_d$, $X_d'$, $X_d''$, $X_q$, $X_q'$, $X_q''$, $T_{d0}'$, $T_{q0}'$, $T_{d0}''$, and $T_{q0}''$ are positive, real-valued parameters. Their definition can be found in \cite[Table 2.1]{hotz2021online}.
The electrical power output of a synchronous machine is given by 
 \begin{equation*} \label{eq:Pe}
    p_e = E_d'' I_d + E_q'' I_q,
 \end{equation*}
 where $I_d$ and $I_q$ can be derived from
 \begin{equation*} \label{eq:i_d_i_q}
    \begin{bmatrix} R & X_d'' \\ -X_d'' & R \end{bmatrix}
    \begin{bmatrix} I_d \\ I_q \end{bmatrix} =
    \begin{bmatrix} E''_d \\ E''_q \end{bmatrix} - 
    \begin{bmatrix} v_d \\ v_q \end{bmatrix}.
 \end{equation*}
Here, $R$ is the armature winding resistance, $v_d = \operatorname{Re}(u)$, and $v_q = \operatorname{Im}(u)$ with the bus voltage $u$.
Finally, the current injected by the synchronous generator is
\begin{equation*} \label{eq:i_inj}
    i_{inj} = -\left(\frac{E_d''}{X_d''} + j\frac{E_q''}{X_q''} \right) e^{j \delta}.
\end{equation*}
%% GOVERNOR %%
\noindent The governors are modeled like in Figure~\ref{fig:governor}. They are driven by the frequency deviation $\Delta \omega$ and the steady-state active power fed
into the network by the corresponding synchronous machine $p_{m0}$. The parameters of the governors are explained in Table~\ref{tab:parameters_governor}.
\begin{figure}[h]
    \center
    \begin{adjustbox}{width=\columnwidth}
    \begin{tikzpicture}
        \draw[stealth-] (-1,0.5) to (-1.5,0.5) node[left] {$R_gp_{m0}$};
        \draw (-0.75,0.5) circle [radius=0.25] node {$+$};
        
        \draw[-stealth] (-0.75,-1) to (-0.75,0.25);
        \draw[-stealth] (-0.5,0.5) to (0,0.5);
        \draw (0,0) rectangle +(1.5,1) node[pos=.5] {$\frac{1}{R_g}$};
        \draw[-stealth] (1.5,0.5) to (2,0.5);
        \draw (2,0) rectangle +(1.5,1) node[pos=.5] {$\frac{1}{1+sT_1}$};
        
        \draw (2.75,0) to (2.55,-0.2) to (2.35,-0.2) node[left, font=\footnotesize] {$V_{min}$};
        \draw (2.75,1) to (2.95,1.2) to (3.15,1.2) node[right, font=\footnotesize] {$V_{max}$};
        
        \draw[-stealth] (3.5,0.5) to (4,0.5);
        \draw (4,0) rectangle +(1.5,1) node[pos=.5] {$\frac{1+sT_2}{1+sT_1}$};
        \draw (5.5,0.5) to (6,0.5);
        \draw[-stealth] (6,0.5) to (6,-0);
        
        \draw (-0.5,-1) to (-1.5,-1) node[left] {$\Delta \omega$};
        
        \draw[-stealth] (-0.5,-1) to (2,-1);
        \draw (2.0,-1.5) rectangle +(1.5,1) node[pos=.5] {$-D_t$};
        \draw (3.5,-1) to (6,-1);
        \draw[-stealth] (6,-1) to (6,-0.5);
        
        \draw (6,-0.25) circle [radius=0.25] node {$+$};
        \draw[-stealth] (6.25,-0.25) to (6.75,-0.25) node[right] {$p_{GOV}$};
    \end{tikzpicture}
    \end{adjustbox}
    \caption{Block diagram of the governor model.}
    \label{fig:governor}
\end{figure}
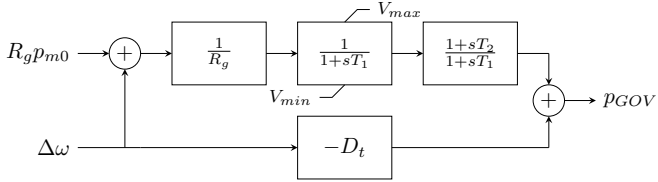
\begin{table}[H]
\caption{Parameters of the governor model.} 
\label{tab:parameters_governor}
\centering
\begin{tabular}{l | l }
 Parameter & Description\\ \hline
 $T_1$ & Governor time constant \\
 $T_2$,  $T_3$ & Turbine time constants \\
 $R_g$ & Turbine governor droop \\
 $D_t$ & Frictional losses factor \\
 $V_{min}$, $V_{max}$ & Valve limits \\
 \end{tabular}
\end{table}
%% EXCITATION SYSTEM %%
\noindent The excitation systems of the synchronous machines are modeled according to the block diagram in Figure~\ref{fig:excitation}. The input $v_{pss}$ comes from the power system stabilizer, $\Delta v$ is the deviation of the bus voltage magnitude $v$ from the voltage set point $v_{OFO}$, and $E_{f0}$ is the steady-state field voltage. The parameters of the excitation system are explained in Table~\ref{tab:parameters_excitation}.

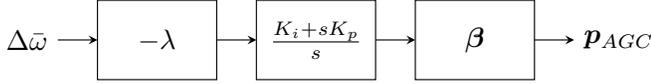
\begin{figure}[h]
    \center
    \begin{adjustbox}{width=\columnwidth}
    \begin{tikzpicture}
        \draw[stealth-] (0,0.5) to (-.5,0.5) node[left] {$\Delta \bar \omega$};
        \draw (0,0) rectangle +(1.5,1) node[pos=.5] {$-\lambda$};
        \draw[-stealth] (1.5,0.5) to (2,0.5);
        \draw (2,0) rectangle +(1.5,1) node[pos=.5] {$\frac{K_i + s K_p}{s}$};
        \draw[-stealth] (3.5,0.5) to (4,0.5);
        \draw (4,0) rectangle +(1.5,1) node[pos=.5] {$\bm\beta$};
        \draw[-stealth] (5.5,0.5) to (6,0.5) node[right] {$\bm p_{AGC}$};
    \end{tikzpicture}
    \end{adjustbox}
    \caption{Block diagram of the automated generation control.}
    \label{fig:AGC}
\end{figure}

\begin{table}[H]
\caption{Parameters of the excitation system model.}
\label{tab:parameters_excitation}
\centering
\begin{tabular}{l | l }
 Parameter & Description\\ \hline
 $K_{ex}$ & Controller gain \\
 $T_a$, $T_b$ & Filter time constants \\
 $T_e$ & Exciter time constant \\
 $E_{min}$, $E_{max}$ & Field voltage limits \\
 \end{tabular}
\end{table}
%% POWER SYSTEM STABILIZER %%
\noindent The power system stabilizer can be seen in Figure~\ref{fig:PSS} and is driven by the frequency deviation $\Delta \omega$. The parameters of the power system stabilizer are explained in Table~\ref{tab:parameters_PSS}. 
\begin{figure}[h]
    \center
    \begin{adjustbox}{width=\columnwidth}
    \begin{tikzpicture}
        \draw[stealth-] (0,0.5) to (-.5,0.5) node[left] {$\Delta \omega$};
        \draw (0,0) rectangle +(1.5,1) node[pos=.5] {$\frac{sK_{PSS}}{1+sT}$};
        \draw[-stealth] (1.5,0.5) to (2,0.5);
        \draw (2,0) rectangle +(1.5,1) node[pos=.5] {$\frac{1+sT_1}{1+sT_3}$};
        \draw[-stealth] (3.5,0.5) to (4,0.5);
        \draw (4,0) rectangle +(1.5,1) node[pos=.5] {$\frac{1+sT_2}{1+sT_4}$};
        \draw[-stealth] (5.5,0.5) to (7.5,0.5) node[right] {$v_{pss}$};
        
        \draw (6.55,-0.2) to (6.35,-0.2) node[left, font=\footnotesize] {$-H_{lim}$};
        \draw (6.95,1.2) to (7.15,1.2) node[right, font=\footnotesize] {$H_{lim}$};
        \draw (6.55,-0.2) to (6.95,1.2);
    \end{tikzpicture}
    \end{adjustbox}
    \caption{Block diagram of the power system stabilizer.}
    \label{fig:PSS}
\end{figure}
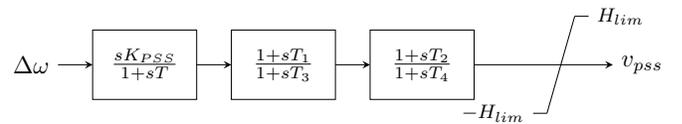
\begin{table}[H]
\caption{Parameters of the power system stabilizer model.}
\label{tab:parameters_PSS}
\centering
\begin{tabular}{l | l }
 Parameter & Description\\ \hline
 $K_{PSS}$ & Controller gain \\
 $T$ & Washout-filter time constant \\
 $T_1$, $T_3$ & Time constants of first lead-lag compensation \\
 $T_2$,  $T_4$ & Time constants of second lead-lag compensation \\
 $H_{lim}$ & Output limit \\
 \end{tabular}
\end{table}
%% AUTOMATED GENERATION CONTROL %%
\noindent The Automated Generation Control can be seen in Figure~\ref{fig:AGC}. It balances the active power generation and consumption in the power system and is driven by the average frequency deviation over all $g$ generators
\begin{equation*} \label{eq:favg}
        \Delta \bar\omega = \frac{\sum_{i \in [1,g]}\Delta\omega_i H_i S_i}{\sum_{i \in [1,g]} H_i S_i}.
\end{equation*}
The vector $\bm{\beta}$ contains the participation factor of each generator and the sum of its elements is 1. The parameters of the Automatic Generation Control are explained in Table~\ref{tab:parameters_AGC}. 
\begin{table}[H]
\caption{Parameters of the automated generation control model.}
\label{tab:parameters_AGC}
\centering
\begin{tabular}{l | l }
 Parameter & Description\\ \hline
 $\lambda$ & Frequency bias factor \\
 $K_p$ & Proportional gain \\
 $K_i$ & Integral gain \\
 $\bm \beta$ & Participation vector
 \end{tabular}
\end{table}
\noindent For more information on the model and the model parameters, the reader is referred to~\cite{hotz2021online}.

\section{Curative Actions via Online Feedback Optimization}\label{sec:OFO}

\begin{figure}
	    \includegraphics[width=\columnwidth]{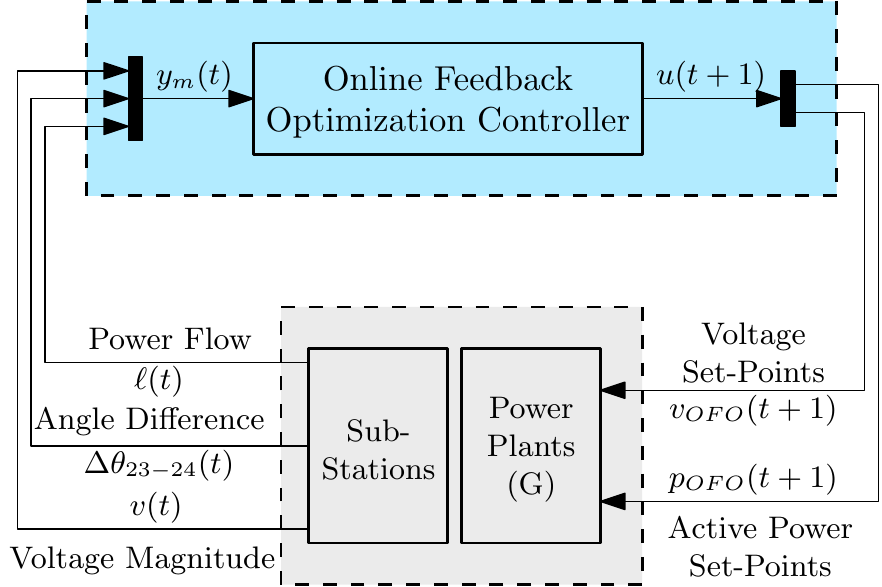}
	    \caption{Schematic representation of the proposed \ac{OFO}-based controller. The controller iteratively gathers power system measurements and updates set-points for the controllable power plants in the grid.}
	    \label{fig:block_diagram}
\end{figure}

The curative actions available in the IEEE 39 bus case are changes in the active power generation set-points and voltage set-points of the generators. Hence, we consider the controllable active power set-points $p_{OFO}$ and voltage set-points $v_{OFO}$ as our input $u=[p_{OFO}^T,v_{OFO}^T]^T$.
We measure the bus voltage magnitudes $v$, all power flows $\ell$, and the phase difference $\Delta\theta_{23-24}$ between buses~23 and~24 and group them in our output $y=[v^T,\ell^T, \Delta\theta_{23-24}]^T$. The block diagram of our controller can be seen in Figure~\ref{fig:block_diagram}.

We encode the goal of reclosing the breaker in the following optimization problem
\begin{equation}\label{eq:opt_problem}
\begin{alignedat}{2}
        \min_u \quad &
        (v_{23}(u)-v_{24}(u))^2 + (\theta_{23}(u) - \theta_{24}(u))^2
        \\
        \textrm{subject to} \quad&
        p_{OFO,\textrm{min}} < p_{OFO} < p_{OFO,\textrm{max}}  
        && \hspace{-9mm}\forall \; \textrm{generators}\\
        & v_{OFO,\textrm{min}} < v_{OFO}<v_{OFO,\textrm{max}}  
        && \hspace{-9mm}\forall \; \textrm{generators}\\
        & v_{\textrm{min}} < v(u) < v_{\textrm{max}} 
        && \hspace{-9mm}\forall \;\textrm{buses}\\
        & \ell_{\textrm{min}} < \ell(u) < \ell_{\textrm{max}} 
        && \hspace{-9mm}\forall \;\textrm{lines}\\
\end{alignedat}
\end{equation}
that minimizes the voltage difference subject to actuator limits and grid constraints. Note, that this optimization problem is specific to this emergency situation, and more work is needed to identify optimization problems for other and more general situations.
As prescribed by the \ac{OFO} approach, we then select an optimization algorithm.
We choose a projected gradient descent algorithm. An \ac{OFO} controller derived from such an algorithm was developed in~\cite{haberle2020non}.
The control update is
\begin{equation} \label{eq:feedbackupdate} 
    u(t+1) = u(t) + \alpha\,{\sigma}_\alpha(u(t),y_m(t))
\end{equation}
with
\begin{align}\label{eq:projection_QP}
\begin{split}
    {\sigma}_\alpha(u,y_m) := \; \arg \min_{w} \; &\left\| w + \nabla_u h(u,y_m) \nabla \Phi (y_m)^T \right\|^2
    \\ 
    \textrm{subject to} \quad &A (u+ \alpha w)\leq b \\ & C (y_m+ \alpha \nabla_u h(u,y_m) w)\leq d \,.
\end{split}
\end{align}
In the controller, $\alpha$ is a gradient step-length (that we set to $\alpha = 3$) and ${\sigma}_\alpha(u,y_m)$ is the projected gradient direction, which is computed via a simple convex quadratic program. Note, that such convex quadratic programs can be efficiently solved for very large numbers of variables and constraints on standard computation hardware.
In this auxiliary optimization program, $\Phi(y_m)$ is the cost function of the optimization problem \eqref{eq:opt_problem} where the output $y$ is replaced by the measurement $y_m$. 
The constants $A$, $b$ $C$, $d$ describe a local linearization of the potentially nonlinear constraints of the optimization problem~\eqref{eq:opt_problem}, see \cite{haberle2020non}.  
The resulting controller determines curative actions based on measurements and the sensitivity $\nabla_u h(u,y_m)$. Overall, we are solving the highly nonlinear and non-convex optimization problem~\eqref{eq:opt_problem} by repeatedly solving the linear and convex problem~\eqref{eq:projection_QP} and utilizing feedback measurements. For an in-depth decision of the convergence of this strategy see~\cite{hauswirth2021optimization}.
$\nabla_u h(u,y_m)$ is the sensitivity matrix of input (set-points) to output (measurements). These are similar to e.g. power transfer distribution factors and we derive them from the steady-state power flow equations, see \cite{bolognani2015fast} for details. We recalculate this sensitivity at every time step, which occurs every 5 seconds. Note, that this sensitivity is calculated many times while solving Optimal Power Flow problems. When solving security-constraint Optimal Power Flow Problems, it is calculated for all considered contingencies. Knowing and calculating $\nabla_u h(u,y_m)$, as needed for our controller, is therefore a reasonable assumption.
Note that, it can also be estimated and real-time adapted from data~\cite{picallo2021adaptive}. Last but not least, the controller is also robust with respect to an inaccurate sensitivity, on which we provide more details in Section~\ref{sec:results}.

\section{Results}
\label{sec:results}

\begin{figure*}[tb]
		\centering
		\begin{tikzpicture}
			\node (image) at (0,0) {%
		\includegraphics[width=1.99\columnwidth]{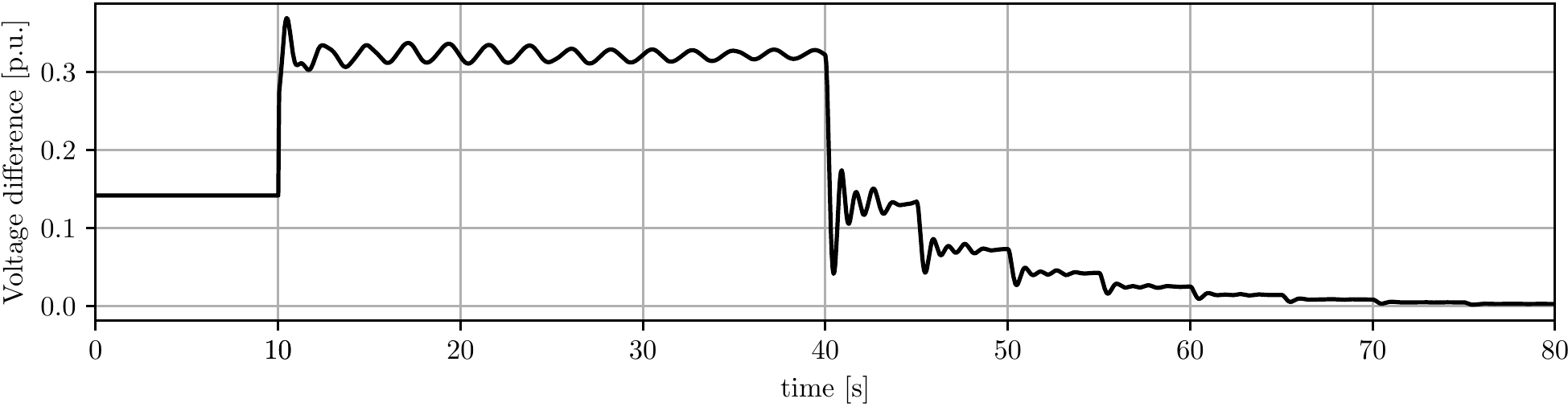}};
   \node[below, text=blue] (t10) at (-5,0.2) {\scriptsize Line trips};
			\node[below, text=blue] (t40) at (1.4,2.1) {\scriptsize \ac{OFO} activated};
		\end{tikzpicture}\\
		\begin{tikzpicture}
			\node (image) at (0,0) {%
				\includegraphics[width=1.99\columnwidth]{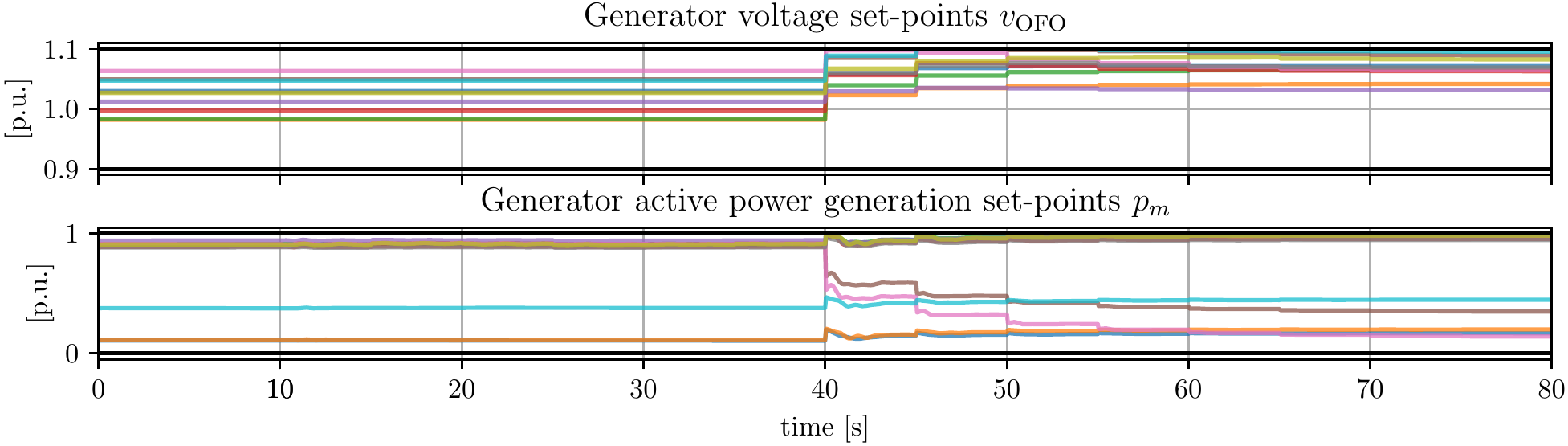}
		};
			%\node[below, text=blue] (t10) at (-2.8,-1.2) {\scriptsize Line trips};
			%\node[below, text=blue] (t40) at (0.2,-1.2) {\scriptsize \ac{OFO} activated};
		\end{tikzpicture}
		\caption{Dynamic power system simulation of a line contingency at 10~seconds and activation of the \ac{OFO} controller at 40~seconds with a sampling time of 5~seconds. $p_m$ is the sum of the \ac{OFO} controller set-point $p_{OFO}$ and the primary and secondary frequency control. The limits for the voltage set-points are 0.9~p.u. and 1.1~p.u. The limits for the active power set-points are 0~p.u. and 1~p.u. Each color corresponds to one generator.}
		\label{fig:voltage_simulation}
	\end{figure*}

The upper panel in Figure~\ref{fig:voltage_simulation} shows the absolute voltage difference between buses~23 and~24. A small absolute voltage difference implies that both the voltage angle difference and voltage magnitude difference between buses~23 and~24 are small, which allows the breaker to be reclosed. The middle panel shows the generators' voltage set-points and the lower panel shows the generators' active power set-points. After~10 seconds, the line connecting the two buses trips and the absolute voltage difference increases. As described in Section~\ref{sec:blackout} local protection might prohibit reclosing the line. Therefore, after~30 seconds, the \ac{OFO} controller is activated to reduce the absolute voltage difference over the breaker.
As can be seen in Figure~\ref{fig:voltage_simulation}, the controller takes effective steps towards minimizing the absolute voltage difference, and within just a few iterations the breaker could be closed again. 
While the controller is minimizing the absolute voltage difference, the constraints in its update law~\eqref{eq:projection_QP} also enforce that the control inputs $u$ are within the actuator capabilities, as can be seen in the lower two panels. Likewise, the constraints on $y$, i.e., bus voltage magnitudes and current flows, can also be enforced.

Overall, the proposed controller quickly reduces the voltage difference. The resulting curative actions include iterative adjustments of the active power and voltage set-points of all generators, showing how complex coordinated interventions may be needed in order to effectively tackle a contingency. 

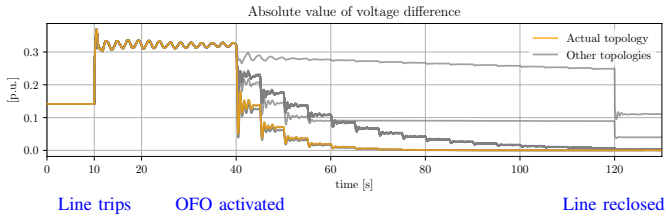
\begin{figure}
	\begin{tikzpicture}
			\node (image) at (0,0) {%
				\includegraphics[width=.99\columnwidth]{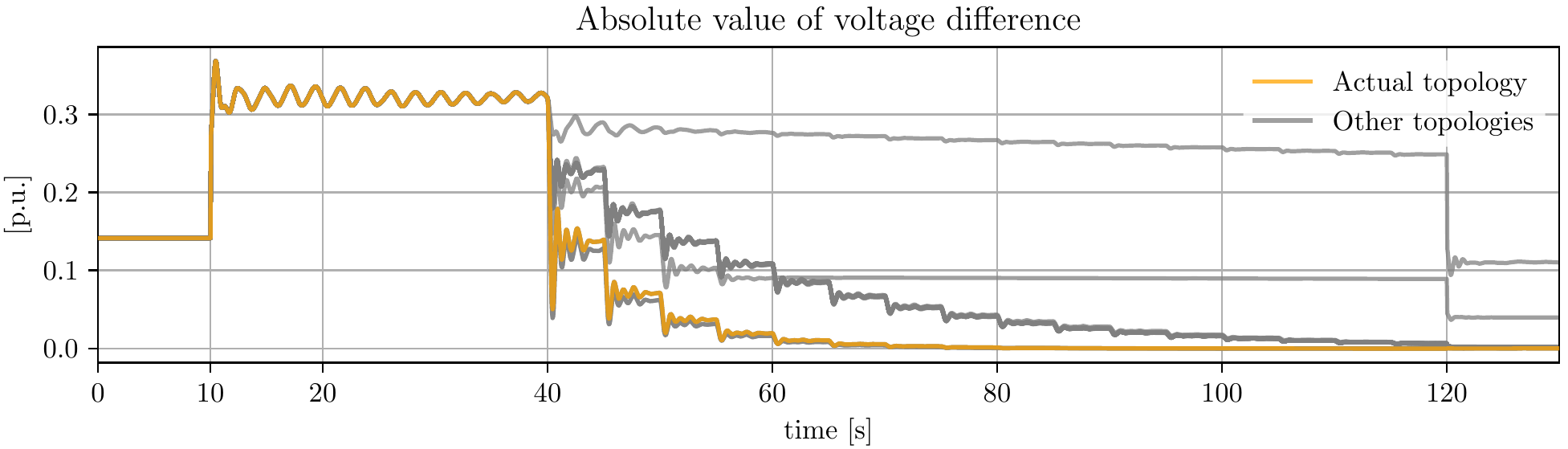}
		};
   \node[below, text=blue] (t10) at (-3.2,-1.2) {\scriptsize Line trips};
			\node[below, text=blue] (t40) at (-1.4,-1.2) {\scriptsize \ac{OFO} activated};
			\node[below, text=blue] (t40) at (3.7,-1.2) {\scriptsize Line reclosed};
		\end{tikzpicture}
		\caption{Behavior of the closed-loop system for several incorrect sensitivities. The line is tripped at 10~seconds, the controller is activated at 40~seconds, and the line is reclosed at 120~seconds.}
		\label{fig:robustness}
	\end{figure}

We also analyze the robustness of our controller against model mismatch.
The only model information used in an \ac{OFO} controller is the sensitivity $\nabla_u h(u,y_m)$.
The sensitivity might be wrong if it was derived in a different operating state or based on a model with wrong parameters or an incorrect topology. In practice, the sensitivity will always have a model mismatch. For our robustness analysis, we calculate the sensitivity based on grid topologies that are different from the topology of our simulation model. More precisely, we derive the sensitivity for a topology where we erased a line from the grid and then use these wrong sensitivities in our controller. In many power grids, the position of switches and breakers is observed, and therefore a model mismatch due to a wrong topology is unlikely to occur. Nevertheless, we choose this source of model mismatch because we consider it to be the most extreme. The results of our robustness analysis can be seen in Figure~\ref{fig:robustness} and show that even with severe model mismatch, the controller is able to reduce the absolute voltage difference and does not become unstable. However, some levels of model mismatch cause very slow performance, and future work should analyze how well the sensitivity needs to be known to guarantee good performance. 
Generally, the robustness against model mismatch is due to the feedback nature of the approach and the fact that our control law~\eqref{eq:feedbackupdate} is an integrator driven by a gradient step, and integral controllers are known to be robust. This robustness was also observed in experiments \cite{ortmann2020experimental} and analyzed mathematically \cite{colombino2019towards}.

Another source of uncertainty that the controller needs to be robust against is that commanded inputs $u$ are not implemented as asked for. For example, the synchronous generators do not follow the commanded input $p_{OFO}$ but the value $p_m$, because the set-points of the governor and the Automatic Generation Control are added on top of $u_{OFO}$, compare Figure~\ref{fig:simulation_model}. The lower panel in Figure~\ref{fig:voltage_simulation} shows the active power generation set-points $p_m$ and one can see that they change continuously and not just every 5~seconds when our controller updates its set-point. Nevertheless, the controller converges because it measures the output $y$ and therefore indirectly the effect of the governor and the Automatic Generation Control.

\section{Conclusion}\label{sec:conclusion}

These preliminary numerical results show that \ac{OFO} controllers have the potential to derive curative actions in real-time after the occurrence of a contingency and to automate some curative actions in emergency power system operations. Such controllers could either be implemented as a decision support tool for the operator or directly as a closed-loop controller.
In our opinion, determining curative actions in real-time is in agreement with the European and North American grid codes, and it definitely reduces the workload in the control room.

Further research is needed to investigate the stability of the interconnection of the controller with the power system dynamics because timescale separation results like those in \cite{timescale-separation} (which assume that grid dynamics are sufficiently faster compared to the rate at which set-points are updated by the controller) turn out to be too conservative for this time-critical application. Furthermore, because we expect the system to work far from nominal operating points during contingencies, robustness to model mismatch needs to be certified for this application more extensively (possibly building on numerical tests like those in \cite{colombino2019towards}).
Last but not least, a broader range of emergency situations needs to be analyzed.

\bibliographystyle{IEEEtran}
\bibliography{IEEEabrv,bibfile}

\end{document}